\documentclass[12pt]{iopart}
\usepackage{iopams}
\usepackage{amssymb,latexsym}
\usepackage{amstext}

\newcommand{\ket}[1]{\vert\mathit{#1}\rangle}
\newcommand{\bra}[1]{\langle\mathit{#1}\vert}
\newcommand{\braket}[2]{\langle\mathit{#1}\vert\mathit{#2}\rangle}

\begin{document}

\title[Mutually unbiased bases]{Group theoretical construction
 of mutually unbiased bases in Hilbert spaces of prime dimensions}
 \author{P \v{S}ulc and  J Tolar}
\address{Department of Physics\\
Faculty of Nuclear Sciences and Physical Engineering         \\
Czech Technical University \\ B\v{r}ehov\'a 7,  CZ - 115 19 Prague,
Czech Republic}
 \ead{sulcp1@km1.fjfi.cvut.cz, jiri.tolar@fjfi.cvut.cz}

 \begin{abstract}
Mutually unbiased bases in Hilbert spaces of finite dimensions are
closely related to the quantal notion of complementarity. An
alternative proof of existence of a maximal collection of $N+1$
mutually unbiased bases in Hilbert spaces of prime dimension $N$ is
given by exploiting the finite Heisenberg group (also called the
Pauli group) and the action of $SL(2,\mathbb{Z}_N)$ on finite phase
space $\mathbb{Z}_N \times \mathbb{Z}_N$ implemented by unitary
operators in the Hilbert space. Crucial for the proof is that, for
prime $N$, $\mathbb{Z}_N$ is also a finite field.
\end{abstract}

 \submitto{J. Phys. A: Math. Theor.}


\section{Introduction}
Position coordinates and corresponding components of linear momentum
of a free quantum particle are --- according to usual quantum
mechanics --- \textit{complementary observables}, a term coined by
Niels Bohr in reference to measurements. The main expression of this
property (for quantum mechanics on the real line) are Heisenberg's
uncertainty relations
 $ \Delta q \Delta p \geq \hbar/2$,  implying that there are no
quantum states with arbitrarily narrow distributions of both
conjugate variables $q$ and $p$. If there were normalized states
with $\Delta q =0$, i.e.~eigenstates of $q$, then $\Delta p$ would
diverge, hence measured values of $p$ couldn't be predicted.
However, due to continuity of spectra of $q$ and $p$, the strict
limits $\Delta q =0$ or $\Delta p =0$ cannot be physically attained.
Complementarity thus means that quantum systems possess properties
that are mutually exclusive: the observation of one of them
precludes the observation of the other. A mathematical expression of
complementarity of coordinate and momentum is
\begin{equation}\label{braket}
\vert \braket{p}{q}\vert^2 = const.,
\end{equation}
i.e.~if we know everything about position, we know nothing of
momentum, and vice versa.

It is interesting and may have deep physical significance that pairs
of complementary observables also exist for systems with
finite-dimensional Hilbert spaces, as noted by J.~Schwinger
\cite{Schwinger*} (see also \cite{Schwinger,Kraus}). For such pairs
precise knowledge of one of them implies that all possible outcomes
of measuring the other one are equally probable, as exemplified in
\eref{braket}.

If complementary observables have non-degenerate spectra,
complementarity can be expressed in terms of the corresponding normalized
eigenstates forming complementary bases. Then a measurement over one
basis provides maximum uncertainty as to the outcome of a
measurement in the other because all $N$ possible outcomes will have
equal probabilities $1/N$. The first attempt to use complementary
bases in quantum state determination is due to Ivanovi\'c
\cite{Ivanovic} who provided explicit formulae for $N+1$ such bases
if $N$ is an odd prime. The idea of using these bases for optimal
quantum state determination was further developed by Wootters
\cite{Woottersprob} and by Wootters and Fields \cite{WoottersFields}
who called them \textit{mutually unbiased}. In the latter paper they
presented a construction of $N+1$ mutually unbiased bases in an
arbitrary prime power-dimensional Hilbert space and also
demonstrated that they may serve as a maximal collection of measurements
for optimal state determination.

 Our main concern are $N$-level systems which provide
basic models proposed for quantum information processing, since
mutually unbiased bases find important applications there
\cite{Nielsen}. Their property that the outcome of a measurement in
one selected basis gives no information about the possible results
of measurements in all other mutually unbiased bases is of advantage
for instance in key distribution protocols in quantum cryptography
\cite{Gisin}. Observables with such a property for two-level systems
(whose vectors are called \textit{qubits} in quantum computation)
are three Pauli matrices. Recently, $d$-level quantum systems (with
$d>2$ and vectors called $qudits$) have come to a closer attention.
It has been shown that such systems can be realized experimentally
and quantum key distribution protocols using qudits have been
introduced (see e.g.~\cite{Alber}). Since such protocols use
mutually unbiased bases in dimensions higher than two, it is
desirable to study constructions of mutually unbiased bases for
higher dimensions, too. It is also known that complementary
observables are useful in quantum state tomography
\cite{Wootters87}.

\section{Complementary observables and mutually unbiased bases}
Mutually unbiased bases in Hilbert spaces of finite dimensions are
closely related to the quantal notion of complementarity. Let us
start with two almost identical definitions.

{\bf Definition 1} \cite{Kraus}. \textit{Two observables
$\mathbf{A}$ and $\mathbf{B}$ of a quantum system with Hilbert space
of finite dimension $N$ are called complementary, if their
eigenvalues are non-degenerate and any two normalized eigenvectors
$\ket{u_i}$ of $\mathbf{A}$ and $\ket{v_j}$ of $\mathbf{B}$ satisfy}
     \begin{equation}
     |\braket{u_i}{v_j}| = \frac{1}{\sqrt{N}}.
    \end{equation}

Then in an eigenstate $\ket{u_i}$ of $\mathbf{A}$ all eigenvalues
$b_1,\ldots,b_N$ of $\mathbf{B}$ are measured with equal
probabilities, and vice versa. This means that exact knowledge of
the measured value of $\mathbf{A}$ implies maximal uncertainty to
any measured value of $\mathbf{B}$. For the next definition note
that the (non-degenerate) eigenvalues $a_i$ of of $\mathbf{A}$ and
$b_j$ of $\mathbf{B}$ are in fact irrelevant, since only the
corresponding orthonormal bases $\ket{u_i}$ and $\ket{v_j}$ are
involved.

{\bf Definition 2} \cite{Woottersprob, WoottersFields}. \textit{Two
orthonormal bases in an $N$-dimensional complex Hilbert space
 \begin{equation*}
 \lbrace \ket{u_i}
|i=1,2,\ldots,N \rbrace \quad \mbox{ and } \quad
\left\lbrace\ket{v_j} |j=1,2,\ldots,N\right\rbrace
 \end{equation*}
are called mutually unbiased if inner products between all possible
pairs of vectors taken from distinct bases have the same magnitude
$1 / \sqrt{N} $,
     \begin{equation}
     |\braket{u_i}{v_j}| = \frac{1}{\sqrt{N}} \quad
\mbox{ for all } \quad i,j \in \left\lbrace 1,2,\ldots,N
\right\rbrace.
    \end{equation}
}

In the above sense one may call two measurements to be mutually
unbiased, if the bases composed of the eigenstates of their
observables (with non-degenerate spectra) are mutually unbiased.
Further, a set of bases is called mutually unbiased if every two
different bases from the set are mutually unbiased.

An important fact was proved in \cite{WoottersFields} (for further
references see also \cite{KlappRott}) that the upper limit to the
maximal possible number of bases that can form a set of mutually
unbiased bases in an $N$-dimensional Hilbert space is $N+1$:

{\bf Theorem 1}. \textit{In an $N$-dimensional Hilbert space, there
cannot be more than $N+1$ mutually unbiased bases.}

Finally we remark that a criterion of equivalence of two pairs of
mutually unbiased bases was formulated in \cite{Kraus}.

\section{Quantum structures in finite-dimensional Hilbert spaces}
Our starting point for quantum mechanics in the Hilbert space of
finite dimension $N$ is a model of quantum kinematics due to H. Weyl
\cite{Weyl}. Its geometric interpretation as the simplest quantum
kinematics on a finite discrete configuration space formed by a
periodic chain of $N$ points was elaborated by J.~Schwinger
\cite{Schwinger}. In \cite{Tolar, StovTolar} we proposed its group
theoretical formulation based on Mackey's system of imprimitivity
\cite{Mackey} which provides a group theoretical generalization of
Heisenberg's commutation relations. For a recent review of the topic
see \cite{Vourdas, Vourdas07}.

In an $N$-dimensional Hilbert space with orthonormal basis
$\mathcal{B} = \left\lbrace\ket{0}, \ket{1}, \ldots
\ket{N-1}\right\rbrace$ we can establish a group generated by
unitary operators $Q_N$, $P_N$ defined by the relations
    \begin{eqnarray}
     Q_N \ket{j} = \omega_N^j \ket{j}, \quad j=0,1,\ldots,N-1, \\
     P_N \ket{j} = \ket{j-1 \pmod{N}};
    \end{eqnarray}
here $\omega_N$ is a primitive $N$-th root of unity, e.g.~$\omega_N
= \exp(2\pi i/N)$. If $\mathcal{B}$ is the standard (or canonical)
basis of $\mathbb{C}^N$, the operators $P_N$ and $Q_N$ are
represented by matrices
    \begin{equation}
    \label{maticeQ}
      Q_N = \mbox{diag}\left(1,\omega_N,\omega_N^2,\cdots,\omega_N^{N-1}\right)
    \end{equation}
and
    \begin{equation}
    \label{maticeP}
     P_N = \left(
    \begin{array}{cccccc}
     0 & 1 & 0& \cdots & 0 & 0 \\
     0 & 0& 1&  \cdots & 0 & 0 \\
     0 & 0 & 0&\cdots & 0 & 0 \\
     \vdots &   & & \ddots &   & \\
     0 & 0 &0 & \cdots & 0 & 1 \\
     1 & 0 &0 &\cdots & 0 & 0
    \end{array} \right)
    \end{equation}
In finite-dimensional quantum mechanics the unitary matrices $Q_N$
and $P_N$ are analogues of exponentials of position and momentum in
the continuous coordinate representation \cite{Weyl}.
Namely, they fulfil an algebraic relation
    \begin{equation}
     \label{qnpnomega}
    P_N Q_N = \omega_N Q_N P_N \\
    \end{equation}
which is analogous to the relation for Weyl's exponential form  of
Heisenberg's commutation relations. Further, $P_N^N = Q_N^N = I_N$,
where $I_N$ is the $N\times N$ unit matrix.

This model has a simple geometric interpretation. The cyclic group
$ \mathbb{Z}_N = \left\lbrace 0,1,\ldots N-1
 \right\rbrace $
serves as a configuration space for $N$-dimensional quantum
mechanics. Elements of $\mathbb{Z}_N$ label the vectors of the basis
$\mathcal{B} = \left\lbrace\ket{0}, \ket{1}, \ldots
\ket{N-1}\right\rbrace$ with the physical interpretation that
$\ket{j}$ is the normalized eigenvector of position at $j \in
\mathbb{Z}_N$. The natural transitive action of $\mathbb{Z}_N $ on
$\mathbb{Z}_N $ via addition modulo $N$ is represented by unitary
operators $U(k) = P_{N}^{k}$. Their action on vectors $\ket{j}$ from
basis $\mathcal{B}$ is given by
  \begin{equation}
    U(k) \ket{j} = P_N^k \ket{j} = \ket{j - k \pmod{N}}
  \end{equation}

The $\mathbb{Z}_N$ analogue of the Fourier transformation
is the discrete Fourier transformation given by the unitary
Sylvester matrix $S_N$ with elements
   \begin{equation}
   \label{sylvestrdefinice}
     (S_{N})_{jk} = \bra{j}S_N
     \ket{k}=\frac{\omega_N^{jk}}{\sqrt{N}}
   \end{equation}
involving powers of $\omega_N$. The relations
   \begin{equation}
   \label{sylvesterPSylevster}
    S_{N}^{-1} P_N S_N = Q_N, \quad S_{N}^{-1} Q_N S_N = P_{N}^{-1}
   \end{equation}
show that the discrete Fourier transform diagonalizes the momentum
operator, i.e.~performs the transition from the coordinate
representation to the momentum representation.

The finite group generated by $\omega_N$, $Q_N$ and $P_N$
     \begin{equation}
     \label{pauligroup}
     \Pi_N = \left\lbrace \omega_N^l Q^i_N P_N^j |
     l,i,j = 0,1,2,\ldots,N-1\right\rbrace
     \end{equation}
consists of $N^3$ unitary matrices and is called the \textit{finite
Heisenberg group} \cite{Balian} or the \textit{Pauli group}
\cite{PatZass}. It has been found useful in connection with mutually
unbiased bases \cite{Ind}. Note also that the set of $N^2$ unitary
matrices $\left\lbrace Q_N^a P_N^b | a,b \in \left\lbrace
0,1,\ldots,N-1 \right\rbrace\right\rbrace$ constitutes,
as Schwinger \cite{Schwinger*} has shown, a complete operator
basis of the Hilbert space of all complex matrices
orthogonal in the sense of the inner product
\begin{eqnarray}
      \mbox{Tr}\left(\left(Q^a_N P^b_N\right)^\dagger
      Q_N^{c} P^d_N\right) = N \delta_{ac}  \delta_{bd}
      \quad \mbox{for all}  \quad  a,b,c,d \in  \mathbb{Z}_N.
\end{eqnarray}

\section{Mutually unbiased bases for prime $N$}
The question whether it is possible to attain the maximal number of
$N+1$ mutually unbiased bases was answered in positive in
\cite{WoottersFields}, but under a number theoretic proviso: a
maximal collection of $N+1$ mutually unbiased bases exists in Hilbert
spaces of dimensions equal to arbitrary powers of prime numbers.
In this paper we will devote our attention to prime dimensions.

In the particular case $N=2$ one finds that the set of eigenvectors
of the Pauli matrices forms a complete collection of three mutually
unbiased bases:
       \begin{equation}\label{dimtwo}
        \left\lbrace \ket{0}, \ket{1} \right\rbrace , \quad
    \left\lbrace \frac{\ket{0} + \ket{1}}{\sqrt{2}},
    \frac{\ket{0} - \ket{1}}{\sqrt{2}}\right\rbrace , \quad
    \left\lbrace  \frac{\ket{0} + i\ket{1}}{\sqrt{2}},
     \frac{\ket{0} - i\ket{1}}{\sqrt{2}}\right\rbrace.
       \end{equation}
The construction of such a set of bases in higher dimensions can be
understood as a generalization of this property. The formulae for
$N+1$ mutually unbiased bases forming a maximal set \textit{for any
odd prime dimension} $N$ were first given (not derived) in
\cite{Ivanovic}; we quote them according to \cite{WoottersFields}:
       \begin{eqnarray*}
        \ket{v_{k}^{(0)}}_j &=& \delta_{j k}, \\
      \ket{v_{k}^{(1)}}_j &=& \frac{1}{\sqrt{N}} e^{\frac{2 \pi i}{N} (j^2 + j k)},\\
       \vdots  & & \\
   \ket{v_{k}^{(r)}}_j &=& \frac{1}{\sqrt{N}} e^{\frac{2 \pi i}{N} (rj^2 + j k)},\\
     \vdots & & \\
     \ket{v_{k}^{(N-1)}}_j &=& \frac{1}{\sqrt{N}} e^{\frac{2 \pi i}{N}((N-1)j^2 + j k)},  \\
      \ket{v_{k}^{(N)}}_j &=& \frac{1}{\sqrt{N}} e^{\frac{2 \pi
      i}{N} j k}.
        \end{eqnarray*}
Here $\ket{v_{k}^{(r)}}_j$ denotes the $j$-th component of the
$k$-th vector in $r$-th basis, $r=0,1,\dots,N$. The first ($r=0$)
basis is the canonical basis, the last one ($r=N$) is its discrete
Fourier transform. Mutual unbiasedness of the bases follows from the
Gauss sums of number theory valid for $p$ odd prime
 \cite{LidlNiederreiter}
    \begin{equation}
    \label{gsuma}
      \left\vert  \sum_{k=0}^{p-1} e^{\frac{2 \pi i}{p}(a k^2 + bk)}
       \right\vert = \frac{1}{\sqrt{p}};
    \end{equation}
here $a,b \in \mathbb{N}$, $a \neq 0$ and $a$ is not an integral
multiple of $p$.

A derivation of these $N+1$ mutually unbiased bases has been given
for any prime dimension $N$ in \cite{Ind} in terms of unitary
operators $Q_N$, $P_N$ defined in Section 3:

{\bf Theorem 2}. \textit{Let $N$ be a prime.
Then the bases composed of eigenvectors of $N+1$ operators
      \begin{equation}
 Q_N,P_N,P_N Q_N, P_N Q_{N}^{2},\dots, P_N Q_{N}^{N-1}
      \end{equation}
are pairwise mutually unbiased and form therefore
a maximal set of $N+1$ mutually unbiased bases.}

In this paper we are going to give an alternative
construction of a complete collection of $N+1$ mutually
unbiased bases in a prime-dimensional Hilbert space by using
the finite phase space related to the finite Heisenberg group.

\section{Finite phase space and its group of automorphisms}
In order to arrive at an independent proof of Theorem 2
we have to introduce the necessary group theoretical notions.

First we need to establish a connection between the finite
Heisenberg group and the \textit{finite phase space} $\Gamma_N =
\mathbb{Z}_N \times \mathbb{Z}_N$, $N=2,3,\dots$ \cite{Balian,Wootters87}.
The elements $(l,i,j)$ of the finite Heisenberg group were given in
\eref{pauligroup} with $l,i,j = 0,1,\ldots,N-1$. Its center
$Z(\Pi_N)$ is the set of those elements of $\Pi_N$ which commute
with all elements in $\Pi_N$,
    \begin{equation}
Z(\Pi_N) = \left\lbrace (l,0,0) |l=0,1,\ldots,N-1\right\rbrace.
    \end{equation}
Since the center is a normal subgroup, we can go over to the
quotient group $\Pi_N / Z(\Pi_N)$. Its elements are the cosets
labeled by pairs $(i,j)$, $i,j = 0,1,\ldots,N-1$. The quotient group
is then identified with the finite phase space $\Gamma_N =
\mathbb{Z}_N \times \mathbb{Z}_N$. To simplify notation, we shall
denote the cosets corresponding to elements $(i,j)$ of the phase
space $ \Gamma_N$ by $Q^i P^j$ without subscripts $N$,
    \begin{equation}
Q^i P^j= \left\lbrace \omega_N^l Q^i_N P_N^j | \quad l =
0,1,\ldots,N-1 \right\rbrace.
    \end{equation}
Note that all operators belonging to the same coset have the same
eigenvectors because they differ only by multipliers $\omega_N^l$.

{\footnotesize
 The definition of the phase space can be concentrated in the exact
 sequence \cite{Balian}
   \begin{equation*}
 1 \rightarrow Z(\Pi_N) \rightarrow \Pi_N \rightarrow
 \Pi_N / Z(\Pi_N) \rightarrow
 \Gamma_N = \mathbb{Z}_N \times \mathbb{Z}_N
 \rightarrow 1
    \end{equation*}
expressing the relation of the projective representation of
 $\mathbb{Z}_N \times \mathbb{Z}_N$ to the central extension by
 $Z(\Pi_N)$ \cite{Weyl}. It is obvious that the correspondence
    \begin{equation*}
\phi : \Pi_N / Z(\Pi_N) \rightarrow \Gamma_N=\mathbb{Z}_N \times
\mathbb{Z}_N \quad : \quad  Q^i P^j \mapsto (i,j),
     \end{equation*}
     is an isomorphism of Abelian groups, since
     \begin{equation*}
     \phi\left(\left(Q^i P^j\right)\left(Q^{i'} P^{j'}\right)\right) =
\phi\left(\left(Q^i P^j\right)\right)\phi\left(\left(Q^{i'}
P^{j'}\right)\right) = \end{equation*}
 \begin{equation*} = (i,j) + (i',j') = (i+i',j+j').
     \end{equation*}
}

 We shall now focus on the group of automorphisms of the phase
space $\Gamma_N$. It was studied in \cite{Balian}. However, we
follow the approach of \cite{Pateraspol}, where instead of cosets
the one-dimensional grading subspaces of the Pauli graded Lie
algebra $gl(N,\mathbb{C})$ were considered and their transformations
under the automorphisms of $gl(N,\mathbb{C})$ were investigated. The
subgroup of inner automorphisms of $gl(N,\mathbb{C})$ was induced by
the action
     \begin{equation}
     \label{automorf}
      \psi_X (A) = X^{-1} A X
    \end{equation}
of matrices $X$ from $GL(N,\mathbb{C})$.

In the same vein we will concentrate on the
automorphisms of the form \eref{automorf}, acting on elements of
$\Pi_N$, which induce permutations of cosets in $\Pi_N / Z(\Pi_N)$.
Since the operators $\omega_N^l Q_N^a P_N^b$ have the same spectra,
the matrices $X$ which induce the automorphisms \eref{automorf} are unitary.
They can be understood as transformation matrices
that transform a unitary operator (of the form $\omega_N^l Q_N^a
P_N^b$) to a different basis, in which the operator is
of the form $ \omega^{m} Q^c P^d$. Thus $X$ is
a transformation matrix between two orthonormal bases.
For explicit forms of matrices $X$ see
\cite{Balian} (for $N$ odd prime), but the results of
\cite{Pateraspol} will suit better to our purpose.

Automorphisms $\psi$ of the form \eref{automorf} are equivalent if they
define the same transformation of cosets in $\Pi_N / Z(\Pi_N)$:
    \begin{equation}\label{condition}
\psi_Y \sim \psi_X \Leftrightarrow Y^{-1} Q^i P^j Y = X^{-1} Q^i P^j
X \quad \mbox{for all}\quad (i,j) \in \mathbb{Z}_N \times
\mathbb{Z}_N.
    \end{equation}
Since the group $\Pi_N / Z(\Pi_N)$ has only two generators --- the
cosets $P$ and $Q$ --- condition \eref{condition} is equivalent to
     \begin{equation}
\psi_Y \sim \psi_X  \Leftrightarrow Y^{-1} P Y = X^{-1} P X \mbox{
and } Y^{-1} Q Y = X^{-1} Q X.
     \end{equation}
If $\psi_Y$ induces a transformation of $\Pi_N / Z(\Pi_N)$, then
there must exist elements $a,b,c,d \in \mathbb{Z}_N$ such that
     \begin{equation}
  Y^{-1} Q Y = Q^a P^b \quad \mbox{and} \quad Y^{-1} P Y = Q^c P^d.
     \end{equation}
It follows that to each equivalence class of automorphisms $\psi_Y$ a
quadruple $(a,b,c,d)$ of elements in $\mathbb{Z}_N$ is assigned.
We shall prove more, namely:

{\bf Theorem 3.} \textit{For $N$ prime there is an isomorphism
$\Phi$ between the set of equivalence classes of automorphisms
$\psi_Y$ and the group $SL(2,\mathbb{Z}_N)$ of $2 \times 2$ matrices
with determinant equal to $1 \mbox{ modulo } N$,
       \begin{equation*}
        \Phi(\psi_Y) = \left( \begin{array}{cc}
        a & b \\
        c & d
       \end{array}\right), \qquad a,b,c,d \in \mathbb{Z}_N;
       \end{equation*}
the action of these automorphisms on $\Pi_N / Z(\Pi_N)$ is given by
the right action of $SL(2,\mathbb{Z}_N)$ on the phase space
$\Gamma_N=\mathbb{Z}_N \times \mathbb{Z}_N$,
       \begin{equation}
        (i',j') =  (i,j) \left( \begin{array}{cc}
        a & b \\
        c & d
       \end{array}  \right).
       \end{equation}
}

\textit{Proof:} To the composition of two automorphisms $\psi_X,
\psi_Y$ corresponding to $(a_X,b_X,c_X,d_X)$ and
$(a_Y,b_Y,c_Y,d_Y)$, respectively, the product of matrices
corresponding to $\psi_X$ and $\psi_Y$ is assigned, as can be
seen from
     \begin{equation*}
(XY)^{-1} Q (XY) = (Y^{-1} Q Y)^{a_X} (Y^{-1} P Y)^{b_X}  =
\end{equation*}
 \begin{equation*}
 =Q^{a_Y a_X} P^{b_Y a_X} Q^{c_Y b_X} P^{d_Y b_X} =
 Q^{a_X a_Y + b_X c_Y} P^{a_X b_Y + b_X d_Y},
     \end{equation*}
and similarly for $P$
      \begin{equation*}
      (XY)^{-1} P (XY) = Q^{c_X a_Y + d_X c_Y} P^{c_X b_Y + d_X
      d_Y}.
     \end{equation*}
Hence
     \begin{equation}
      \Phi(\psi_X \psi_Y) = \Phi(\psi_X) \Phi(\psi_Y)
     \end{equation}
 and $\Phi$ is an injective homomorphism.

Now matrix elements $a,b,c,d$ cannot be chosen
arbitrarily. Consider the action of $\psi_Y$:
     \begin{eqnarray}
     \label{aqa}
       Y^{-1} Q Y = Q^{a} P^{b}\quad \Longrightarrow \quad Y^{-1} Q_N Y &=&
       \mu Q_N^a P_N^b, \quad |\mu|=1, \\
      \label{apa}
       Y^{-1} P Y = Q^{c} P^{d}\quad \Longrightarrow \quad Y^{-1} P_N Y &=&
       \lambda Q_N^c P_N^d, \quad |\lambda|=1.
      \end{eqnarray}
By multiplying equation \eref{aqa} by equation \eref{apa} once from
the left and once from the right, we obtain
     \begin{equation}
  P_N Q_N Y =  \mu \lambda Y Q_N^{c} P_N^{d} Q_N^{a} P_N^{b},\quad
      Q_N P_N Y  = \mu \lambda Y Q_N^{a} P_N^{b} Q_N^{c} P_N^{d} .
     \end{equation}
Using the commutation relation \eref{qnpnomega} we obtain
      \begin{equation}
       \omega_N^{-ad} \mu \lambda Y Q_N^{a+c} P_N^{b+d}= P_N Q_N Y
       = \omega_N^{-1} Q_N P_N Y = \omega_N^{-1} \omega_N^{-bc}
       \mu \lambda Y Q_N^{a+c} P_N^{b+d}
      \end{equation}
leading to the condition
       \begin{equation}
        \omega_N^{-ad} = \omega_N^{-bc - 1}.
       \end{equation}
It will be fulfilled if and only if $ad - bc = 1 \pmod{N}$, i.e.
       \begin{equation}
         \mbox{det}\left(  \begin{array}{cc}
        a & b \\
        c & d
       \end{array}
         \right) = 1 \pmod{N}.
       \end{equation}
This means that to every $\psi_Y$ acting on $\Pi_N / Z(\Pi_N)$ a
matrix from $SL(2,\mathbb{Z}_N)$
       \begin{equation*}
        \Phi(\psi_Y) = \left( \begin{array}{cc}
        a & b \\
        c & d
       \end{array}
         \right)
       \end{equation*}
is assigned. Now to every coset from $\Pi_N / Z(\Pi_N)$ an element
$(i,j)$ of the phase space $\mathbb{Z}_N \times \mathbb{Z}_N$ was
associated. So finally we check that the action of $\psi_Y$ on $Q^i
P^j$ is given by
       \begin{equation*}
       (i',j') =  Q^{i'}P^{j'} = \psi_Y\left(Q^{i}P^{j}\right) =
       Y^{-1}Q^i P^j Y =
       \end{equation*}
       \begin{equation*}
       =Y^{-1} Q^i Y Y^{-1}  P^j Y  =
       Q^{ia + jc} P^{ib + jd} = (ia+jc, ib+jd),
       \end{equation*}
and this means that the transformation of $(i,j)$ can be written
as the right action of $SL(2,\mathbb{Z}_N)$ on $\mathbb{Z}_N \times
\mathbb{Z}_N$
       \begin{equation}
        (i',j') =  (i,j) \left( \begin{array}{cc}
        a & b \\
        c & d
       \end{array}    \right).
              \end{equation}
Finally observe that mapping $\Phi$ is an isomorphism, since
       \begin{equation*}
\Phi : \psi_X \mapsto I_2 \quad \Longrightarrow \quad
      \psi_X \in [\psi_{I_N}]
             \end{equation*}
follows from
$$
  (i',j')=(i,j)\quad \mbox{if and only if} \quad X^{-1}Q^i P^j X =
    I_{N}^{-1}Q^i P^j I_N.  \quad \Box
$$

We conclude this section with

{\bf Lemma 1} \cite{Novotny}. \textit{The right action of $\mbox{SL}(2,\mathbb{Z}_N)$
on the phase space $\mathbb{Z}_N \times \mathbb{Z}_N$ does not
change the determinant of a matrix composed of components of two
vectors from $\mathbb{Z}_N \times \mathbb{Z}_N$.}

    {\it Proof}:
     Consider two vectors $(i,j)$ and $(k,l)$ from $\mathbb{Z}_N \times
\mathbb{Z}_N$ and the matrix
    $   \left( \begin{array}{cc}
        i & j \\
        k & l
       \end{array}
         \right).  $
Since the action of $A$ is given by $(i',j') = (i,j)A$ and $(k',l')
= (k,l)A$, and because $\mbox{det} \, A = 1$, one immediately gets
the result
    \begin{equation*}
    \mbox{det}\left( \begin{array}{cc}
        i' & j' \\
        k' & l'
       \end{array}
         \right) = \mbox{det}\left( \left( \begin{array}{cc}
        i & j \\
        k & l
       \end{array} \right) A
         \right)  =
\end{equation*}
    \begin{equation}\label{det}
         =\mbox{det} \left( \begin{array}{cc}
        i & j \\
        k & l
       \end{array} \right)  \mbox{det} \, A
        = \mbox{det}\left( \begin{array}{cc}
        i & j \\
        k & l
       \end{array}
         \right). \quad \Box
    \end{equation}

{\bf Remarks}. For $N$ prime the right action of
$\mbox{SL}(2,\mathbb{Z}_N)$ on the phase space
$\mathbb{Z}_N \times \mathbb{Z}_N$ has exactly two orbits ---
the single point $\{(0,0)\}$ and
$\mathcal{O}_N=\mathbb{Z}_N \times \mathbb{Z}_N \backslash \lbrace
(0,0) \rbrace$
consisting of $N^{2}-1$ points. The stationary subgroup
of the point $(1,0)$ from $\mathcal{O}_N$ is the Abelian
subgroup $ \lbrace \left( \begin{array}{cc}
        1 & 0 \\
        b & 1
       \end{array}
         \right)  \vert b=0,1,\dots,N-1     \rbrace $
of order $N$.
Hence the order of $\mbox{SL}(2,\mathbb{Z}_N)$ is $N(N^{2}-1)$.
Further, according to Lemma 1, the determinant \eref{det} is
an invariant of the right action of $\mbox{SL}(2,\mathbb{Z}_N)$
on $\Gamma_N \times \Gamma_N$.
Let us note that $\mbox{SL}(2,\mathbb{Z}_N)$ transformations
of the finite phase space were also studied in \cite{Vourdas96}.

\section{New construction of the maximal set of mutually unbiased
bases for $N$ prime}
In this section the finite phase space and its transformations
of the form \eref{automorf} will be used to introduce an interesting
algebraic structure that proves the existence of $N+1$
mutually unbiased bases for prime $N$, thus providing an alternative
approach to their construction. We shall exploit the fact that,
for prime $N$, $\mathbb{Z}_N$ is a finite field, i.e., there is also
a multiplicative group structure modulo $N$ in $\mathbb{Z}_{N}^*=
\mathbb{Z}_N \backslash \lbrace 0 \rbrace$.

Our construction starts with the partition of the finite phase
space $\Gamma_N = \mathbb{Z}_N \times \mathbb{Z}_N$ into equivalence
classes $[(i,j)]$ defined by the equivalence relation \textit{
$(i,j) \sim (i',j')$, if there exists $r \in \mathbb{Z}_{N}^*$
such that $(i',j') = (ri,rj)$, where the multiplication is
understood modulo $N$.}

We exclude the trivial class $[(0,0)]$ containing only $(0,0)$.
Then the orbit \textit{$\mathcal{O}_N=
\mathbb{Z}_N \times \mathbb{Z}_N \backslash \lbrace
(0,0) \rbrace$ is decomposed into $N+1$ classes $[(1,0)]$ and
$[(i,1)]$ where $i = 0,1,\ldots,N-1$}. The fact that $\mathbb{Z}_N$
is a field for prime $N$ is crucial in the proof that every element
of the orbit $\mathcal{O}_N$ belongs to some class.  Since each class has $N-1$
elements, this decomposition contains $N^2 -1$ elements in total,
with the only element $(0,0)$ not included.

If an element is of the form $(0,i)$ or $(i,0)$,
then it is obvious that it belongs to classes
$[(0,1)]$ or $[(1,0)]$, respectively. An element of the form $(i,j)$,
$i,j = 1,2,\ldots,N-1$, will belong to the class $[(k,1)]$ where $k \in
\left\lbrace 1,2,\ldots,N-1 \right\rbrace$ is the solution of $kj =
i \pmod{N}$. The existence and uniqueness of such $k$ is guaranteed
by the fact that $\mathbb{Z}_N$ is a field for $N$ prime.
The partition into classes can be visualized in the following
 table of all elements in $\mathbb{Z}_N
\times \mathbb{Z}_N \backslash \lbrace (0,0) \rbrace$:
       \begin{center}
\renewcommand{\arraystretch}{1.9}
\begin{tabular}{|l|lllll|}
\hline
  & 0 & 1 & 2 & $\cdots$ & N-1 \\
  \hline
0 &  & \fbox{(0,1)} & (0,2) & $\cdots$  & (0,N-1) \\
1 & \fbox{(1,0)} & \fbox{(1,1)} & (1,2) & $\cdots$ & (1,N-1)  \\
2 & (2,0) & \fbox{(2,1)} & (2,2)  & $\cdots$  &  (2,N-1) \\
$\vdots$ & $\vdots$  & $\vdots$  & $\vdots$ & $\ddots$     &  $\vdots$ \\
N-1 & (N-1,0) & \fbox{(N-1,1)} & (N-1,2)& $\cdots$ & (N-1,N-1) \\
\hline
\end{tabular}
\end{center}

Here every element $(i,j)$ corresponds to a coset $Q^i P^j$.
All operators in the same coset differ just by a complex
multiplier. Every multiple $(ri,rj)$ of a vector $(i,j)$ by $ r \in
\left\lbrace1,2,\ldots,N-1\right\rbrace $ will therefore correspond
to the coset $Q^{ri} P^{rj}$. Because of relation \eref{qnpnomega}
it is obvious that operators $(Q^i P^j)^r$ and $(Q^{ri}P^{rj})$
belong to the same coset.
An important consequence is that elements $(ri,rj)$,
$r = 1,2,\dots, N-1$, correspond to commuting operators, hence have the
same eigenvectors. Thus we have proved

{\bf Lemma 2}. \textit{If $N$ is a prime, then there are exactly $N+1$
classes of elements from $\mathcal{O}_N=\mathbb{Z}_N \times \mathbb{Z}_N
\backslash \lbrace (0,0) \rbrace$, each class containing $N-1$
elements. All elements of the same class correspond to
commuting operators with the same eigenvectors.}

We will now demonstrate that the bases composed of the eigenvectors
of two different operators
corresponding to elements from distinct classes in $\mathcal{O}_N$
are mutually unbiased.

  {\bf Theorem 4}. \textit{
Let $N$ be a prime and let $(a,b)$ and $(c,d)$ be two elements
from $\mathcal{O}_N=\mathbb{Z}_N \times
\mathbb{Z}_N \backslash \lbrace (0,0) \rbrace$ which
belong to distinct classes $[(a,b)]\neq [(c,d)]$. Then the bases
composed of eigenvectors of the operators from the corresponding
cosets $Q^a P^b$ and $Q^c P^d$ are mutually unbiased.}

  {\it Proof}:
The first step is to show that the bases composed of
eigenvectors of $Q_N$ and $P_N$ are mutually unbiased.
 This follows directly from equation
\eref{sylvesterPSylevster}. Namely,
  \begin{equation}
   \label{sqvejevlastnivektor}
      P_N S_N \ket{j}  = S_N Q_N \ket{j}
      = \omega_{N}^{j}S_N \ket{j},
  \end{equation}
where $ \ket{j} $ is an eigenvector of $Q_N$,
so $S_N \ket{j}$ is an eigenvector of $P_N$.
Further, because of \eref{sylvestrdefinice},
the inner product of $\ket{j}$ and $S_N \ket{k}$ has
absolute value
  \begin{equation}\label{complementarity}
   \left\vert(\ket{j}, S_N \ket{k})\right\vert =
   \left\vert\bra{j}S_N \ket{k}\right\vert=
   \left\vert \frac{\omega_{N}^{jk}}{\sqrt{N}} \right\vert =
   \frac{1}{\sqrt{N}}.
  \end{equation}
Hence if we have two elements where one belongs to the class $[(1,0)]$ and
the other to the class $[(0,1)]$, then we already know that
their corresponding bases are mutually unbiased, because they
are composed of eigenvectors of $Q_N$ and $P_N$, respectively.

Because of the partition of $\mathcal{O}_N$ it is
sufficient to consider now only the case of two distinct elements
$(a,1)$ and $(b,1)$, with  $a,b \in
    \left\lbrace 1,2,\ldots,N-1\right\rbrace$, $a \neq b$.
We are going to show that the bases of eigenvectors of the corresponding
operators $Q_N^a P_N$ and $Q_N^b P_N$ are mutually unbiased (hence
also the eigenvectors of powers of these operators).
According to Theorem 3, to unitary operators $X$ that permute
the cosets in the Heisenberg group
    \begin{equation*}
      X^{-1} Q^{i} P^{j} X = Q^{i'} P^{j'}
    \end{equation*}
matrices from $\mbox{SL}(2,\mathbb{Z}_N)$ are assigned.
Conversely, to every matrix from $\mbox{SL}(2,\mathbb{Z}_N)$
there is an equivalence class of unitary operators which induce
the same permutation of the cosets.
In this sense a special unitary representation of
$\mbox{SL}(2,\mathbb{Z}_N)$ was described in \cite{Balian}.

We will now show, if $a \neq b$, then there exists a
matrix $A$ from $\mbox{SL}(2,\mathbb{Z}_N)$ such that
    \begin{equation*}
      (a,1) A = (\tilde{a},0) \quad \mbox{and} \quad (b,1) A = (0,\tilde{b})
    \end{equation*}
If we indeed can find such a matrix, then there exists a
corresponding unitary operator $X$ such that
    \begin{equation*}
      X^{-1} Q^a P X =  Q^{\tilde{a}}, \qquad
      X^{-1} Q^b P X =  P^{\tilde{b}} ,
     \end{equation*}
hence the eigenvectors of $Q^a P$ and $Q^b P$ can be expressed as
$X \ket{j}$ and $X S_N \ket{k}$, respectively.
According to \eref{complementarity} modulus of their inner product is
     \begin{equation*}
           | (X\ket{j},X S_N \ket{k})| =
      |(\ket{j},S_N \ket{k})| = \frac{1}{\sqrt{N}},
     \end{equation*}
proving that these bases are mutually unbiased, too.

To prove the existence of a unique matrix $A \in
\mbox{SL}(2,\mathbb{Z}_N)$ with the desired properties we apply
Lemma 1 implying
    \begin{equation*}
     \mbox{det}\left( \begin{array}{cc}
        a & 1 \\
        b & 1
       \end{array}
         \right) = a - b \pmod{N} = \mbox{det}\left( \begin{array}{cc}
        \tilde{a} & 0 \\
        0 & \tilde{b}
       \end{array}
         \right) = \tilde{a} \tilde{b} \pmod{N}
     \end{equation*}
and we select $\tilde{a},\tilde{b} \in \mathbb{Z}_N$ such that
$ \tilde{a} \tilde{b} = a- b \pmod{N}$. Equivalently we look
for a matrix $C = A^{-1}\in \mbox{SL}(2,\mathbb{Z}_N)$
producing the inverse transformation
    \begin{eqnarray*}
     (\tilde{a},0) C = (\tilde{a},0) \left( \begin{array}{cc}
        \alpha & \beta \\
        \gamma & \delta
       \end{array}
         \right)  = (a,1) ,\\
       (0,\tilde{b}) C = (0,\tilde{b}) \left( \begin{array}{cc}
        \alpha & \beta \\
        \gamma & \delta
       \end{array}
         \right)  = (b,1).
    \end{eqnarray*}
This gives us the following equations to compute the elements of
$C$:
    \begin{eqnarray}
     \label{alfabeta}
      \tilde{a} \beta &=& 1 \pmod{N},\\
      \label{aalpha}
      \tilde{a} \alpha &=& a \pmod{N},\\
      \label{bgamma}
      \tilde{b} \gamma &=& b \pmod{N}, \\
       \label{bdelta}
      \tilde{b} \delta &=& 1 \pmod{N}.
    \end{eqnarray}
The fact that $N$ is a prime guarantees that each of these equations
has unique solution in $\mathbb{Z}_N$. Having the values of entries
$\alpha,\beta,\gamma,\delta$, we still need to check that
$C= \left( \begin{array}{cc}
        \alpha & \beta \\
        \gamma & \delta
       \end{array}
         \right)$ belongs to $\mbox{SL}(2,\mathbb{Z}_N)$.
By multiplying equations \eref{aalpha} and \eref{bdelta} and
subtracting the product of \eref{bgamma} and \eref{alfabeta} we
obtain
      \begin{equation*}
       \tilde{a} \tilde{b} (\alpha \delta - \beta \gamma) = a - b \pmod{N}.
      \end{equation*}
Since $\tilde{a} \tilde{b} = a - b \pmod{N}$ we have
      \begin{equation*}
      \mbox{det} \ C = \alpha \delta - \beta \gamma =  1 \pmod{N},
      \end{equation*}
verifying that $C$ indeed belongs to $\mbox{SL}(2,\mathbb{Z}_N)$.
The inverse matrix
$    A =    C^{-1} = \left( \begin{array}{cc}
        \delta & - \beta \\
        - \gamma & \alpha
       \end{array}
         \right)$
will then transform pairs $(a,1)$ and $(b,1)$ into $(\tilde{a},0)$ and
$(\tilde{b},0)$, respectively:
      \begin{equation*}
        (a,1) C^{-1} = (\tilde{a},0), \quad
    (b,1) C^{-1} =(0,\tilde{b}).
      \end{equation*}
To complete the proof, it is easy to see that for pairs $(b,1)$,
$(1,0)$ and $(b,1)$, $(0,1)$, $b=1,2,\ldots,N-1$, there exist
unique transformation matrices from $\mbox{SL}(2,\mathbb{Z}_N)$
such that
      \begin{equation*}
       (b,1) A_{1}(b) = (0,1), \; (1,0) A_{1}(b) = (1,0) \;
       \Longrightarrow \quad
  A_{1}(b) = \left( \begin{array}{cc}
        1 & 0 \\
        - b & 1
       \end{array}
         \right)
      \end{equation*}
and
      \begin{equation*}
       (b,1) A_{2}(b) = (b,0), \; (0,1) A_{2}(b) = (0,1) \;
       \Longrightarrow \quad
A_{2}(b) = \left( \begin{array}{cc}
        1 & b^{-1} \\
        0 & 1
       \end{array}
       \right).
      \end{equation*}
Hence pairs of  bases composed of eigenvectors of pairs of operators
$Q^b P$, $Q$ and $Q^b P$, $P$ are mutually unbiased. Thus we have
shown that there exist $N+1$ mutually unbiased bases in a Hilbert
space of prime dimension $N$. We have therefore reached the same
conclusion as \cite{Ind}. In our case the mutually unbiased bases
are composed of eigenvectors of operators
 \begin{equation}\label{operators}
  Q_N, P_N, Q_N P_N, Q_N^2 P_N, \dots, Q^{N-1}_N P_N,
  \end{equation}
while the operators in Theorem 2 are only modified using
 \eref{qnpnomega}.

Note that one could make a different choice of representatives
of the classes forming the partition: we could have alternatively
used e.g.~the pairs $(1,a),$ $a=1,2,\dots,N-1$, instead of $(a,1)$,
and the mutually unbiased bases would be
given by bases composed of eigenvectors of operators
 \begin{equation*}
  Q_N, P_N, Q_N P_N, Q_N P_N^2, \dots, Q_N P_N^{N-1}.
 \end{equation*}

To provide a constructive proof we should give an explicit way
 to construct the bases out of the canonical basis $\mathcal{B}$.
 Let us denote the bases composed of eigenvectors
 of \eref{operators} by
 $$ \mathcal{B}=\mathcal{B}_{(1,0)},\; \mathcal{B}_{(0,1)},\; \mathcal{B}_{(1,1)},
\; \mathcal{B}_{(2,1)},\; \dots,\; \mathcal{B}_{(N-1,1)}.
 $$
We know that the map $\mathcal{B}_{(1,0)}\rightarrow\mathcal{B}_{(0,1)}$
is implemented by the unitary operator $S_N$. The next step
$\mathcal{B}_{(0,1)}\rightarrow\mathcal{B}_{(1,1)}$, leaving
$\mathcal{B}_{(1,0)}$ intact, clearly corresponds to the above transformation
matrix $A_{1}(-1)$ from $\mbox{SL}(2,\mathbb{Z}_N)$. Its iterations
will generate further steps.
A unitary transformation $D_N$ which implements $A_{1}(-1)$,
$$
  D_{N}^{-1}Q_N D_N = Q_N, \qquad
  D_{N}^{-1}P_N D_N = \varepsilon_N^{-1} Q_N P_N,
$$
can be taken from \cite{Pateraspol}; it is diagonal,
$$
D_N = \mbox{diag} \ (d_0,d_1,\dots,d_{N-1}), \quad
 d_j = \varepsilon_{N}^{-j}\omega_{N}^{{j}\choose {2}},
$$
where $\varepsilon_N=1$ if $N$ is odd, $\varepsilon_N=
\sqrt{\omega_N}$ if $N$ is even.
In this way we arrive at a sequence of unitary maps
\begin{equation*}
 \mathcal{B}_{(1,0)}  \stackrel{S_N}{\rightarrow}
 \mathcal{B}_{(0,1)} \stackrel{D_N}{\rightarrow}
\mathcal{B}_{(1,1)}\stackrel{D_N}{\rightarrow}
\mathcal{B}_{(2,1)} \stackrel{D_N}{\rightarrow}
\dots  \stackrel{D_N}{\rightarrow}\mathcal{B}_{(N-1,1)},
 \end{equation*}
and the composite unitary operators $D_{N}^b S_N$, $b=0,1,\dots,N-1$
will produce all the bases starting from the canonical one.
$\quad \Box$

{\bf Example} $N=2$. The phase space $\Gamma_2$ consists of 4
elements $(0,0)$, $(1,0)$, $(0,1)$, $(1,1)$. The group
$\mbox{SL}(2,\mathbb{Z}_2)$ (also known as the group of invertible
$2\times 2$ matrices over the simplest finite field $F_2$) with 6
elements
$$
\left( \begin{array}{cc} 1 & 0 \\  0 & 1 \end{array} \right),
\left( \begin{array}{cc} 0 & 1 \\  1 & 0 \end{array} \right),
\left( \begin{array}{cc} 1 & 0 \\  1 & 1 \end{array} \right),
\left( \begin{array}{cc} 1 & 1 \\  0 & 1 \end{array} \right),
\left( \begin{array}{cc} 1 & 1 \\  1 & 0 \end{array} \right),
\left( \begin{array}{cc} 0 & 1 \\  1 & 1 \end{array} \right),
$$
acts transitively on the orbit
$\left\lbrace (1,0),(0,1),(1,1)\right\rbrace$.
Unitary operators transforming the bases \eref{dimtwo}
\begin{equation*}
\mathcal{B}_{(1,0)}\quad  \stackrel{S_2}{\rightarrow}
\quad \mathcal{B}_{(0,1)}\quad \stackrel{D_2}{\rightarrow}
\quad \mathcal{B}_{(1,1)}
\end{equation*}
are
$$
S_2 =\frac{1}{\sqrt{2}}
\left( \begin{array}{cc} 1 & 1 \\ 1 & -1 \end{array} \right),
\quad
D_2 =\left( \begin{array}{cc} 1 & 0 \\  0 & -i \end{array} \right).
$$

\section{Concluding remarks}
The question whether the maximal number $N+1$ can be attained for a
composite dimension $N$, where $N$ is not prime nor a power of a
prime, still remains an open problem. The answer is
not known even for the simplest case $N=6$ where $N+1=7$. A simple
argument \cite{Ind} leads to $3$ such bases. Some numerical attempts
to find further mutually unbiased bases were not successful. So it
remains unclear whether it is indeed possible to reach the maximal
number $N+1$ of them for the Hilbert space of dimension $6$ and in
other composite dimensions as well.

Although the relation between the eigenvectors of $Q_N^i P_N^j$ and
mutually unbiased bases was observed e.g.~in \cite{Ind}, the
relation between the decomposition of the phase space $\mathbb{Z}_N
\times \mathbb{Z}_N$ whose elements correspond to cosets in the
finite Heisenberg group and the existence of mutually unbiased bases
has been left unnoticed so far. We were thus able to give an
independent constructive proof of Theorem 2 using group theory.
However, the proof heavily depends on properties that are a
consequence of $N$ being a prime. The extension of our proof to the
case of Galois fields with prime powers $N$ will be subject of a
future publication. It might also be interesting to investigate
whether our procedure would provide better insight in the problem of
existence or non-existence of the maximal number of mutually
unbiased bases in composite dimensions.

\section*{Acknowledgements}
J.T. is indebted to S. Stenholm for turning his attention to this
interesting and highly topical problem, and to E. Pelantov\'a for
critical reading of the manuscript. J.T. also thanks J. Patera and
Centre de Recherches Math\'ematiques, Universit\'e de Montr\'eal for
hospitality. Partial support by the Ministry of Education of Czech
Republic (projects MSM6840770039 and LC06002) is gratefully
acknowledged. Thanks are due to one of the referees who suggested
several improvements of the text.

\end{document}